\documentclass[12pt,a4paper]{article}
\usepackage[utf8]{inputenc}
\usepackage{amsmath}
\usepackage{amsfonts}
\usepackage{amssymb}
\usepackage{graphicx}
\usepackage{authblk}

\begin{document}
\baselineskip=16pt

\title{Confinement and Deconfinement\\
in\\ Gauge Theories: A Quantum Field Theory\footnote{{\it I dedicate this article as homage to my revered teacher, Professor Alladi Ramakrishnan.}}}
\author{A. P. Balachandran}
\affil{Physics Department, Syracuse University, NY 13244-1130, USA}
\date{}
\maketitle

\begin{abstract}
After a brief recount of small and large gauge transformations
and the nature of observables, we discuss superselection sectors in gauge
theories. There are an infinity of them, classified by large gauge
transformations. Gauge theory sectors are labelled by the eigenvalues of a
complete commuting set (CCS) of these transformations.\\

~~~In QED, the standard chemical potential is one such operator generating
global U(1). There are many more given by the moments of the electric field
on the sphere at infinity. In QCD, the CCS are constructed from the two
commuting generators spanning a Cartan subalgebra.\\

~~~Large gauge transformations commute with the Hamiltonian and preserve the equations of motion. They form an infinite number of `classical symmetries'. But most of them are anamolous changing the superselection sectors. \\

~~~We show that any element of a large gauge transformation can be
added to the standard Hamiltonian as a generalised chemical potential without changing
field equations and that in QCD, they lead to confined and deconfined
phases . A speculation about the physical meaning of these chemical
potentials is also made.\\



\end{abstract}

\section*{Introduction}
~~We propose a new mechanism for confinement-deconfinement
transitions in gauge theories. It is based on a generalised chemical potential in
the Hamiltonian which was discussed briefly by Balachandran et al.\cite{BNPRV} and which is associated with `large gauge transformations’. It defines both superselection sectors, and in non-abelian gauge theories, their
evolution as well and suggests several new phenomena.

~~Let us first review the basics of gauge transformations
as formulated by our group and reviewed in \cite{Andres_Bal}. In later sections, we will develop the announced results.

~~The spacetime unless otherwise stated is 3+1 dimensional Minkowski
$M^{3,1}$ and the metric will be  $(1,-1,-1,-1)$ diag. The gauge group $\mathcal{G}$ is the group of maps from spacetime to a compact Lie group $G$. On a spatial slice, if $g$ belongs to $\mathcal{G}$, then $g(\vec{x})$
approaches an element $h(\hat{x} )$ of $G$ as $|\vec{x} |$  goes to infinity \cite{Bal_Vaidya}.

~~The limit where the limit $|\vec{x}| \rightarrow \infty$ of $g(\vec{x})$ does not exist is not considered here. 

~~The Lie algebra-valued connection will be denoted by $A_\mu$ and its conjugate
electric field will be $E^\mu$. If $\lambda_\alpha$ form a ( hermitean )
basis for the Lie algebra of $G$ with the normalisation
$\mathrm{tr} \lambda_\alpha \lambda_\beta = 2 \delta _{\alpha, \beta}$
in a defining representation, an $N \times N$ one for $SU(N)$,
then we can write $A_\mu = A_\mu^\alpha \lambda_\alpha$ and
$E_\mu = E_\mu^\alpha \lambda_\alpha$. For $U(1)$, we can put
electric charge for $\lambda_\alpha$.

~~In a gauge theory, an infinitesimal gauge transformation can be written on a
spatial slice as
\begin{equation}
    Q(\Xi)=\int_0^{\infty} d^3x \langle \left( D_i  E_i+ J_0 \right) \Xi \rangle
\end{equation}
where\\

~~ a) $\langle \cdots \rangle $  indicates trace over Lie algebra
elements and $D_i$ is covariant derivative. In the $U(1)$ case, there is
no need for trace and $D_i$ becomes ordinary derivative $\partial_i$.

~~b) $\Xi$ is a Lie algebra-valued test function. In the non-abelian case,
we can write it at a point $\vec{x}$  as
$\Xi^\alpha (\vec{x}) \lambda_\alpha$ where the functions $\Xi^\alpha$ approach functions of
$\hat{\vec{x}} =\vec{x} / |\vec{x} |$ as $| \vec{x}|$ goes to infinity. The test function can be any smooth function so that all local gauge transformations can be generated.

~~If all $\Xi^\alpha$ go to zero at infinity adequately fast , we can partially
integrate without surface terms and write
\begin{equation}
    Q(\Xi) = \int_{-\infty}^{+\infty} d^3x \langle   ( -E_i D_i \Xi  +  J_0 \Xi) \rangle
\end{equation}
and that is the standard Gauss law operator. So it vanishes on the quantum states
and maps to zero in the GNS construction.

~~But if $\Xi$ do not become zero functions at infinity, $Q(\Xi)$ need not
vanish on quantum states. They generate the Sky group of Balachandran and
Vaidya \cite{Bal_Vaidya}.  In either case, regardless of the asymptotic behaviour of
$\Xi$, they commute with the algebra of all  observables $\mathcal{A}$ due
to locality. 

~~We have argued for this fact many times elsewhere \cite{Andres_Bal}. This result has many consequences as we 
outline below.

\section*{Novel Generalised Chemical Potentials}
~~Let $H_0$ be the standard full gauge theory Hamiltonian
without any sort of chemical potential  which generates equations of motion.
Pick a $\Xi$ and a constant $\mu$ with energy dimension
and consider the Hamiltonian
\begin{equation}
    H = H_0 + \mu Q(\Xi) := H_0 + H_1.
\end{equation}
~~The new term will not affect equations of motion as it commutes with all
local observables. It will change the fields  in equations of motion
only by local gauge transformations which will not affect
equations of motion. 

~~$Q(\Xi)$ may be called an infinite number of `classical symmetries'. But most are anomalous as we later discuss. We will call $Q(\Xi)$ the {\it generalised} non-abelian chemical potential. 

~~Non-abelian chemical potentials have been discussed in previous work on finite temperature field theories \cite{FTFT}.

 ~~Later we will interpret $Q(\Xi)$ as the coupling of the experimental setup with the system and suggest a scheme to resolve the observation problem in quantum physics. 

~~For $U(1)$ and constant $\Xi$, $Q(\Xi)$ becomes
the standard chemical potential.  

~~The constant $\mu$ sets the scale at which $Q(\Xi)$ becomes significant.

~~There is no point in making $\mu$ functions of $\vec{x}$ as $\Xi$
are already spatial functions.

~~The exponentials of $Q(\Xi)$ for different test functions
generate the Sky group $\mathcal{G}$ and its group algebra $\hat{\mathcal{G}}$ \cite{Bal_Vaidya}.

~~For non-abelian target group $G$ such as $SU(3)$, this algebra is
non-commutative since
\begin{equation}
    [ Q(\Xi_1), Q (\Xi_2) ] = Q ([\Xi_1, \Xi_2] ).
\end{equation}
 
~~ As Dirac taught us, an irreducible representation of $\mathcal{A}$ is 
characterised by a vector state diagonal in a complete
commuting set (CCS). Among them is a vector state diagonal in a CCS from $\hat{\cal{G}}$. This vector state may or may not be preserved under the
time evolution induced by $\mu Q(\Xi) \in H$ ($H_0$ will not affect it).
We will interpret the former as a colour deconfining state and the latter
as a (generically) colour-confining state for reasons given below. In the latter case,
for generic situations, the orbit of the vector state is ergodic as we shall
also see below.

~~The term $Q(\Xi )$ is already present in the Hamiltonian when it is
derived from the gauge theory Lagrangian by Legendre transformation,
as was observed in an earlier paper \cite{BNPRV}.
It is the term 
\begin{equation}
    \int_{-\infty}^{+\infty} d^3x < (- D_i A_0 E_i + A_0 J_0 ) >
\end{equation}
where $\mathrm{tr} ~\langle \cdots \rangle$ is over the Lie algebra indices as usual  and $J_0$ is the Lie algebra valued charge density. It is usually discarded by treating
$A_0$ as a classical field vanishing fast towards infinity and
doing a partial integration. Then it becomes Gauss’s law and hence
zero as an operator.

~~But $A_0$ need not vanish at infinity. It has become $\Xi$ in the current 
notation. 

~~ The extra term in $H$ will not spoil the commutativity
of spatial translation operators with $H$. But with $H$ as $P_0$ and with the usual
unaltered spatial  translation generators, $P_\mu +\mu Q( \Xi ) \delta_{\mu,0}$
 will not transform as a four vector under Lorentz transformations : Lorentz
 symmetry is broken at the operator level But as the extra term does not affect
equations of motion, the latter will maintain their covariance.

~~ However, it has long been known that Lorentz invariance is broken by infrared effects and gauge transformations \cite{FMS, Bal_QCD}, so that there is no need for concern about Lorentz breaking
from this chemical potential.

\section*{Dynamics from $H$}
~~This is determined by the  expectation values in the vector states of the observables. The latter
in turn are classified by the eigenvalues of the superselection 
operators for the vector states. In a particular superselection sector,
a complete set of commuting superselection operators (CCS) are diagonal.

~~The CCS is an abelian subalgebra, preferably a maximal abelian
subalgebra, of the Sky algebra with generators ${ Q(\Xi)}$.
For example, in QED, the conventional choice for $Q(\Xi)$ is the
multiple of the electric charge operator $Q_N$. The test function $\Xi$ in this case
goes to a constant on the celestial sphere $S^2$ ( with coordinates 
$\hat{x}$).

~~But it can also go to any smooth function of $\hat{x}$. Then the
superselected operators become infinite dimensional.

~~In QCD and in non-abelian gauge theories, CCS is much richer. 
First we choose a CCS from the enveloping algebra of $SU(3)$ Lie algebra.
That is spanned by its quadratic and cubic Casimir operators , a
Casimir of an `isospin’ $SU(2) \in SU(3)$ , a `third component ‘ $I_3$ 
of its isospin and the hypercharge Y commuting with $I_3$. Then in
addition to the above Casimirs, we can diagonalise the commuting
Sky operators $Q(\tilde{\Xi})$ where $\tilde{\Xi}$ is a linear
combination $I_3$ and $Y$ with coefficients becoming functions of 
$\hat{x}$ at infinity.

~~For now, let us consider the case where the coefficient functions are
constants at infinity. Then the superselection sector is labelled by $I_3$ and $Y$ 
in an irreducible representation ( IRR) of $SU(3)$ such 
as a triplet or octet. But a generic $\mu Q( \Xi )=H_1$ will not commute
with $Q(\tilde{\Xi})$ and will change the superselection sector. 
Local observables and evolution by the standard $H_0$ will preserve it, 
but not so the chemical potential.

~~This is the new feature in QCD : a generalised non-abelian chemical potential may not preserve the superselection sector. We will discuss the orbit of the superselection sector below, indicate its ergodic features and argue that the mean values of coloured observables in this situation are all $0$. This result substitutes for colour confinement. But the expectation values of Casimir operators of $G$ are constants of motion for $H$ and can certainly be measured. Hence the representation of the coloured sector, if it is $\mathbf{3}, \mathbf{3*}, \mathbf{8},\cdots$ can be determined.

~~But we can also choose a $\Xi$ commuting with $\tilde{\Xi}$, the diagonalised
CCS. It will
affect evolution by an $\hat{x}$-dependent phase, but will not change the
superselection sector at all. We can call such states as characterising
colour deconfined phases.

~~We need explicit examples to which we now turn.

\section*{ The Colour-Confined and Deconfined Phases}
~~The title is misleading. Presumably in the popular literature, confinement is supposed to mean that coloured
states are not in the domain of the Hamiltonian. It is not clear if numerical work based on Wilson lines accomplish such a result.

~~What emerges from our analysis leads to a related result :
the mean values of coloured observables averaged over the interaction 
time are 0, in this phase. All the same, as Casimirs are colour singlets, they can be
determined as previously remarked.

~~Here is a worked ergodic example for $SU(3)$. The superselection
sector is one with $\lambda_{3,8}$ ( and Casimirs)  diagonal and in the
triplet representation. Their 
joint eigenvalues are (1,1),(-1,1) and (0,-2). Any one of them gives 
a density matrix on the coloured observables. Their time evolution is
given by the chemical potential, assumed time independent. 

~~If the picked vector state is an eigenstate of the chemical potential, it will be preserved in time, and coloured expectation values too will be preserved in time : these are the sectors with colour deconfinement.

~~So let us pick a generalised chemical potential which does not preserve them, say
\begin{equation}\label{eq:5}
  H_1 = Q(\Xi ), \Xi = \mu_1 \lambda_1 + \mu_8 \sqrt 3 \lambda _8, \qquad
\mu_1/\mu_8   \mathrm{~~~irrational}.  
\end{equation}

~~The $\mu$’s have energy dimension which sets the scale of this term.
What sets this scale ? It seems to be the same effect which sets the
scale of the standard chemical potential in
finite temperature field theories. 

~~This operator evolves colour (but not local observables) and is our $H_1$.

~~In this example, the two terms in $H$ commute, making computations easy.
But $|(0, 2) \rangle$ is an eigenstate of $H_1$, so ergodicity can show up only
in the remaining states. But that is fine to illustrate the phenomenon.

~~A calculation shows that
\begin{equation}
    e^{it \Xi} = e^{it \mu_1  \lambda_1}  e^{i t \mu_8 \sqrt{3} \lambda_8},
\end{equation}
which gives the action of $e^{it H_1}$ on a chosen vector state.
 
~~The vector ( 0,0,1) in the basis indicated above is an eigenstate of
$H_I$ and defines a deconfined phase. $H_1$ has eigenvalue $-2\mu_8$ and $e^{itH_1}$ has
periodicity $\pi/\mu_8$. But under time evolution, the other two vectors never 
come back to the starting value as claimed.

~~That is because $\frac{\mu_1}{\mu_8}$ is irrational by choice. So too are the ratios of periods. 

~~If the ratios are rational, the orbit of vector states are periodic. 

~~The strong interaction
time is about $10^{-23}$ secs. No experiment can probe the evolution of the system during this time. 
What is observed is the average of an operator during this time. But such an average is 
expected to be zero, especially if the scales $\mu_i$ are large compared to QCD scales. 
 
~~For the vector state, we can also choose an eigenstate of $H_1$
such as the one given by the choice $(0,0,1)$. It changes
just by a phase under time evolution and hence the corresponding 
density matrix does not change at all. Expectation values of colured
observables are then time-independent and there is no theoretical
issue in observing them. So we call them colour-deconfined phases.

\section*{Open Question}
~~It is reasonable to ask if the colour-deconfined phases are attractors
for the confined ones. That will involve perturbing them and observing
if they will relax back to the deconfined phase.

~~But in quantum theory, a unitary perturbation of short duration seems
inadequate for this task. After it is switched off, the system will keep
evolving unitarily after the switch-off time from wherever it finds itself.
A perturbation with a POVM may help.  

~~A better formulation
of the question seems needed. 

\section*{An Interpretation of the Chemical Potential}
~~One supposes that it has an interpretation similar to the standard one:
$\lambda X N$, with $N$ = number operator and $\lambda$ = a constant. In QED, it is the generator of
$U(1)$ gauge transformations so that it commutes with all observables,
adding it to the QED Hamiltonian $H_0$ as the $H_1$ above will not
affect equations of motion. It is like the abelian $Q(\Xi)$ of QED
discussed above. It is in fact what one gets for the choice 
$\Xi (x ) = \Lambda$ for all $x$.

~~When $\Xi$ has an $\hat{x}$ dependent limit for $|\vec{x}|$ going
to $\infty$, $Q(\Xi)$ generates the Sky group \cite{Bal_Vaidya} yielding more generalised chemical potentials. One can consider $\Xi$ approaching definite combinations of spherical harmonics at infinity getting any number of novel chemical potentials. Just like the phase diagram of temperature 
$T$ versus $\lambda$, one can also consider multidimensional phase plots.

~~In the non-abelian case, such a generalised chemical potential can change 
the non-abelian coloured sector which is partially labelled by the eigenvalues of a CCS basis from the Lie algebra. 
For $H_1$ can be any element of the Sky group and need not commute with
the diagonalised elements. So in general it will generate an ergodic orbit in the space of 
states as discussed above. 

~~Note that the evolution is via superselection sectors. It is like the
problem studied by \cite{Asorey_1, Facchi}.


~~Elsewhere we  have argued that quantum observations are done by 
coupling  experimental operators which form an \textit{abelian} algebra \cite{Bal_Tomography}.
We next show that the superselection sectors naturally generate an abelian
algebra  and that our states from CCS are on this abelian algebra. It is this abelian
 algebra that is observed.
 
~~When $H_1$ is switched off, the superselection sector no longer evolves and
the expectation values of the abelian algebra must be giving information
about the system.  The suggestion here is similar to that of Fiorini and
Immirzi, Wightman \cite{WFCollapse} and probably others like Bohr and Landsman.
 
~~Therefore  we can suggest that $H_1$ is the evolution Hamiltonian for
 superselection sectors and reflects the coupling of the experimental
 apparatus to the system. 
 
~~ Thus let us suppose that the experiment is 
initiated at time $t=0$. At this time, assuming that the state is mixed in colour 
with zero mean value for coloured observables,
which  the experiment knows,  $H_1$ makes its appearance in $H$.
After a time $T$, when the experiment ceases,  and $H_1$  becomes $0$,
the evolved mixture of superselection sectors no longer further evolves. Colour singlets 
can then be measured and  the evolution of their  expectation values
from $0$ to $T$ should reveal information about the operator $H_0$ and 
what has happened to the original state.

~~It has been assumed here that the original state is mixed in colour.
If it is pure for colour  and evolves at time T to a pure state say, that too should reveal features of the underlying dynamics.  

~~We have to show the emergence of a commutative algebra from 
superselection operators.  The conjectures regarding this emergence now follow.

\section*{The Commutative Algebra at Infinity in QED}
~~ Let us first recall how to create a charge $q$ state localised at 
\begin{equation}
    e_\infty = \lim_{\tau\rightarrow\infty}~\tau e
\end{equation}  
in QED from the vacuum vector  $| \Omega \rangle$,
assumed to be invariant under all gauge transformations \cite{mrs}.

~~Let $W(x, e)$ be the Wilson line from $\vec{x}$  to infinity 
along the direction $e$ : 
\begin{equation}
    W(x,e) = exp\left\{iq\int~d\tau e^{\lambda}A_{\lambda}(x+e\tau)\right\}=exp(iq \phi(x,e))
\end{equation}
in terms of the `escort' field $\phi$.

~~Let $\psi (x)$ be any charge $q$ field.   Then
\begin{equation}
    exp(iq \phi(x,e)) \psi (x) | \Omega \rangle
\end{equation}
has no local charges and a charge $q$ blip at $e_\infty$. That is
the case for any $\vec{x}$.

~~If $\pi_{q,e} (\mathcal{A})$ is the charge $q$ representation of
the algebra  $\mathcal{A}$ of local observables realised on (1), 
it acts on (1) by left multiplication.

~~  The vector states $H_{q, e_\infty }$ in this representation of $\mathcal{A}$ are
\begin{equation}\label{eq:11}
    \pi_{q,e} (\mathcal{A}) exp(i q \phi(x,e) ) \psi(x)| \Omega \rangle.
\end{equation}
~~Under a large gauge transformation $exp( iq \Lambda)$, 
every vector in this sector transforms with the  fixed phase  $exp(i q \Lambda 
( e_\infty ))$  where  this has been 
defined as
\begin{equation}
    exp(iq \lim_{\tau \rightarrow \infty} \Lambda(x + \tau e ) ). 
\end{equation}
~~So, this charge $q$ superselection sector is labelled by the
representation $Q(\Xi)$ where $\Xi$ has the value
\begin{equation}
    \Xi_\infty = \lim_{\tau\rightarrow\infty}~\Xi(x+\tau e).
\end{equation}


~~ For  
\begin{equation}\label{eq:2}
    exp (iq \phi(x,e')) \psi (x) | \Omega \rangle , \qquad e' \neq e,
\end{equation}
 since  there are certainly $\Xi$ with different values at $e$ 
 and $e'$,  no small gauge transformation or local observable can
 map  the superselection sector defined by \eqref{eq:11} to that defined by
\eqref{eq:2}. 

~~Hence
\begin{equation}\label{eq:3}
    \pi_{q,e} ( (a) exp(iq \phi(x,e)) \psi (x) |\Omega \rangle
\end{equation}
     
\begin{equation}\label{eq:4}
    \pi_{q,e'}( (b) exp(iq \phi(x,e') \psi (x) | \Omega \rangle
\end{equation}
 for any $a, b$ in $\cal{A}$ are eigenvectors with different 
eigenvalues  for such $Q(\Xi)$ and hence
orthogonal. The Hilbert spaces $ H_{q, e,e'}$ built on  
\eqref{eq:3} and \eqref{eq:4} are orthogonal.

   

~~  Let $P_{q,e}$ be the  `projection' operator  for  $H_{q,e}$ ,
like $|x \rangle  \langle x |$ for position in quantum mechanics.
Then
\begin{equation}\label{eq:5a}
    P_{q,e}  P_{q,e'} =0  \qquad \mathrm{if}  ~~e \neq e', \qquad
\end{equation}

~~The set of $e$'s form a de Sitter space.

~~Hence
\begin{equation}\label{eq:6}
    RHS \mathrm{~ of ~\eqref{eq:5a}} = \delta_{e',e}   P_{q,e} 
\end{equation}
  where $\delta_{e,e'}$ is the Dirac delta function on de Sitter space  defined by
\begin{equation}\label{eq:7}
    \int de ~f(e) \delta(e', e)  = f(e')
\end{equation}
where $de$ is the Lorentz invariant volume on the de Sitter space and $f$ is any smooth function thereon.

~~It follows that \
\begin{equation}\label{eq:8}
    I (f ) =  \int de ~f(e) P_{q,e}
\end{equation}
obeys the abelian algebra
\begin{equation}\label{eq:9}
    I(f) I(g) = I(f\,g)
\end{equation}
 where $f\,g$ is  defined by the pointwise multiplication of $f$ 
 and $g$.
    
    ~~With $*$ as complex conjugation and with \textit{sup norm}, 
we get a commutative C*-algebra $\mathcal{C}$ with spectrum as the de Sitter space. 
Elements of this algebra label the superselection sectors associated with a charge $q$ sector. 

\section*{The Commutative Algebra at Infinity  in QCD}

~~ The remarks above can be adapted to any non-abelian gauge theory. But there is an additional feature. At each $e$, for $SU(2)$ say, we can diagonalise $Q(\mathbb{I}\sigma_3)$, $\mathbb{I}$ as in what follows = constant function on $\mathbb{R}^3$, or its entire orbit under $SU(2)$, the Grassmanian $Gr(1,2)$. For $SU(3)$, it is $Gr(2,3)$ etc.  

~~ A paper `Holography in Gauge Theories' is being drafted where these issues are again discussed.

\section*{Epilogue}
There are many issues that remain open and to be addressed. 

\section*{Acknowledgement}
 I have benefitted by discussions with  Manolo Asorey, Bruno Carneiro da Cunha,
 Arshad Momen,  Paraneswaran Nair, Sasha Pinzul, Amilcar Queiroz, Babar Qureshi and Sachin Vaidya. Much of this work was done at the Institute of Mathematical Sciences. I thank my hosts Ravidran, Sanatan Digal and other colleagues there for their wonderful hospitality. It would have been impossible to complete this paper without help from Arshad Momen and Pramod Padmanabhan. I thank them.


\end{document}